\documentclass[aps,prl,twocolumn,superscriptaddress,letterpaper]{revtex4-2}

\usepackage{graphicx}
\usepackage{bm}
\usepackage{amssymb}
\usepackage{color}
\usepackage{amsmath}
\usepackage{physics}
\usepackage[colorlinks=true,linkcolor=blue,anchorcolor=black,citecolor=blue,filecolor=black,menucolor=black,runcolor=black,urlcolor=blue]{hyperref}

\begin{document}

\title{Magnon-induced phononic Chern insulator}

\author{Rui-Chang Shen}
\affiliation{Department of Physics, The Chinese University of Hong Kong, Shatin, Hong Kong SAR, China}

\author{Yihao Yang}
\email{yangyihao@zju.edu.cn}
 \affiliation{State Key Laboratory of Extreme Photonics and Instrumentation, ZJU-Hangzhou Global Scientific and Technological Innovation Center, Zhejiang University, Hangzhou 310027, China.}
  \affiliation{The Electromagnetics Academy at Zhejiang University, College of Information Science and Electronic Engineering, Zhejiang University, Hangzhou 310027, China.}
   \affiliation{Zhejiang Key Lab. of Intelligent Electromagnetic Control and Advanced Electronic Integration, Jinhua Institute of Zhejiang University, Zhejiang University, Jinhua 321099, China.}

\author{Haoran Xue}
\email{haoranxue@cuhk.edu.hk}
\affiliation{Department of Physics, The Chinese University of Hong Kong, Shatin, Hong Kong SAR, China}
\affiliation{State Key Laboratory of Quantum Information Technologies and Materials, The Chinese University of Hong Kong, Shatin, Hong Kong SAR, China}

\begin{abstract}

High-frequency artificial phononic crystals offer a low-loss platform compatible with on-chip integration, yet realizing Chern phononic phases at GHz frequencies remains challenging. Here, we propose a magnon-induced phononic Chern insulator in a honeycomb phononic crystal hybridized with ferromagnetic islands at the hexagon centers. A circularly polarized Kittel mode couples to the surrounding phonons with a phase winding, which breaks time-reversal symmetry and opens a full Chern gap. In the large-detuning regime, this mechanism leads to an effective Haldane-type phononic model with magnon-induced complex hopping. By tuning the magnon-phonon interaction, the full hybrid system accesses Chern phases with tunable Chern numbers $|C|=1$ and $|C|=2$. The predicted gaps can exceed realistic phonon and magnon linewidths, enabling their observation in GHz acoustic devices. Our work establishes chiral magnon--phonon hybridization as a route to magnetically reconfigurable topological phononics.

\end{abstract}

\maketitle

\textit{Introduction---}
Phonons, the quantized excitations of lattice vibrations, are fundamental carriers of thermal energy, mechanical signals, and coherent information~\cite{Maldovan2013}. The introduction of topological band theory into acoustic and elastic systems has given rise to topological phononics, enabling robust boundary transport and unconventional control of sound and mechanical waves~\cite{Wang2015,mousavi2015topologically,yang2015topological,khanikaev2015topologically,ni2015topologically,Susstrunk2015,nash2015topological,kane2014topological,xiao2015geometric,He2016,Yan2018,ma2019topological,xue2022topological}. Most experimental realizations, however, operate at kilohertz-to-megahertz frequencies, where the comparatively long acoustic wavelengths limit miniaturization and integration. Gigahertz (GHz) phonons, by contrast, combine compact wavelengths and low dissipation with compatibility with integrated electromechanical architectures~\cite{Hatanaka2014,delsing20192019}. They can furthermore interface coherently with superconducting qubits,
spin excitations, microwave-frequency resonators, and ferromagnetic dynamics~\cite{Gustafsson2014,Chu2017,Satzinger2018,safavi2019controlling,Weiler2011}, making them promising information carriers for hybrid quantum and classical devices. To date, GHz topological-phononics experiments have
primarily realized valley-Hall, crystalline, or pumping-based phases~\cite{Zhang2022,zhang2024near,Xu2025}. Although such phases can suppress scattering under appropriate symmetry conditions, they cannot provide the strictly one-way boundary transport of a Chern insulator and thus remain vulnerable to symmetry-breaking disorders~\cite{leykam2026limitations}.

Realizing a phononic Chern phase requires breaking the time-reversal symmetry and opening a complete bulk gap. Existing approaches have employed gyroscopic motion~\cite{Wang2015,nash2015topological}, circulating fluids~\cite{yang2015topological,khanikaev2015topologically,ni2015topologically,souslov2017topological,ding2019experimental}, Floquet modulation~\cite{fleury2016floquet,darabi2020reconfigurable}, and charge‑doped piezomagnetic platforms~\cite{zhang2023general}. These schemes have established the basic principles of acoustic Chern transport, but their reliance on moving media, continuous external driving, or material-specific responses presents substantial challenges for miniaturized high-frequency devices. 

Magnons, the collective excitations of spin precession in ordered magnets, provide a natural microscopic source of time-reversal symmetry breaking. Their intrinsic GHz-frequency dynamics and magnetic tunability have made them important carriers of coherent information~\cite{Philipp2021,rameshti2022cavity,yuan2022quantum}. Magnons couple to lattice vibrations through effects like magnetoelastic and magneto-rotation interactions~\cite{Kittel1958,Xu2020}, and suitable symmetry breaking can convert these interactions into nonreciprocal acoustic responses~\cite{Xu2020,Shah2020,Liao2023,Liao2024}. 

\begin{figure*}
\centering
\includegraphics[width=0.9\textwidth]{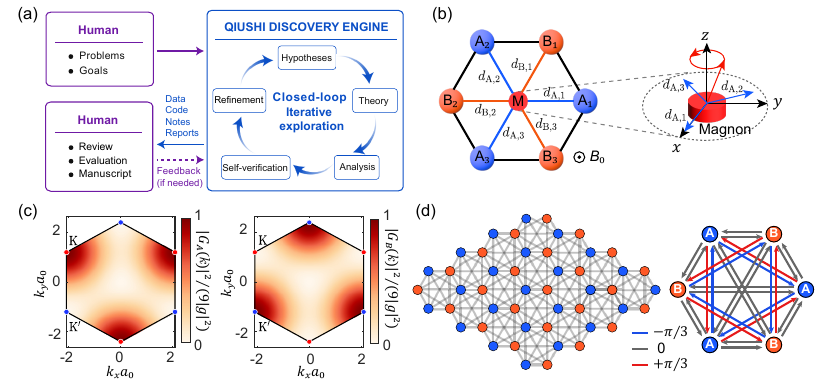}
\caption{(a) Schematic of Human-Agentic hybrid workflow. (b) A honeycomb phononic crystal with one phonon mode on each A and B sublattice site. A magnetic island at the plaquette center supports a right-handed circular Kittel mode. The six surrounding phononic sites are labeled in their physical counterclockwise order as $\mathrm A_1,\mathrm B_1,\mathrm A_2,\mathrm B_2, \mathrm A_3,\mathrm B_3$. Their radial vectors are $\bm d_{\alpha,i}=a_0(\cos\theta_{\alpha,i}, \sin\theta_{\alpha,i})$, with $\theta_{\mathrm A,i}=2\pi(i-1)/3$, $\theta_{\mathrm B,i}=(2i-1)\pi/3$, and $a_0$ is the distance from the plaquette center to each surrounding phononic site. (c)  Momentum-space intensity of the magnon--phonon terms $|G_{\rm A }(\bm k)|^2$ and $|G_{\rm B }(\bm k)|^2$, showing that the magnon couples selectively to sublattice A at $\rm K$ and to sublattice B at $\rm K'$.   (d) Schematic of the magnon-induced effective phonon hopping, showing $H_{\rm eff}- H_{\rm ph}$, where only the magnon-induced coupling between phonons is depicted. The coupling phase is illustrated in the right panel. Each NN bond receives virtual-magnon contributions from the center magnons of its two adjacent plaquettes. }
\label{fig1}
\end{figure*}

Here, we propose a magnon-induced phononic Chern insulator in a honeycomb phononic crystal containing ferromagnetic islands. Unlike previous studies of topological magnon-phonon polaritons~\cite{Takahashi2016,Go2019,Thingstad2019,Zhang2019,zhang2020}, our scheme incorporates magnon--phonon coupling into an artificial phononic-crystal platform, where magnons not merely serve as a constituent of a hybrid excitation, but as virtual mediators that imprint a synthetic gauge flux on a lattice and thereby produce a globally gapped phononic Chern phase. Specifically, due to the discrete rotational symmetry, the magnon couples to the surrounding phonons with a phase winding. This induces virtual complex phonon hoppings and an effective Haldane-type flux, opening a complete Chern gap with a tunable Chern number up to two. We also present the phase diagram of the model and analyse the gap size using realistic parameters, further paving the way for possible experiments.

\textit{Human--AI discovery workflow}---
The theoretical framework was developed through an AI scientific-discovery workflow centered on the Qiushi Discovery Engine~\cite{yang2026end}, as illustrated in Fig.~\ref{fig1}(a). Starting from the open-ended objective of using magnons to realize a phononic Chern insulator, the engine autonomously constructed the initial lattice model, derived its effective Hamiltonian, explored candidate topological mechanisms, and performed closed-loop consistency checks and self-verification. Human researchers subsequently examined the physical assumptions, independently verified the analytical and numerical results, and formulated the final interpretation. This human--AI workflow therefore combined autonomous theoretical exploration with human scientific validation.

\begin{figure*}
\centering
\includegraphics[width=\textwidth]{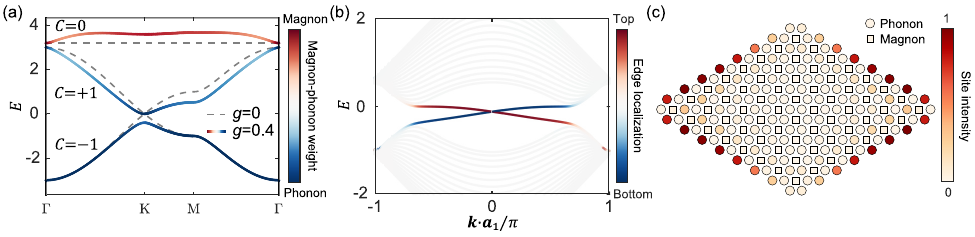}
\caption{
(a) Bulk band structures in the absence (gray dashed curves) and presence (colored solid curves) of the magnon-phonon coupling, which explicitly opens a gap at the Dirac points. The lower, middle and upper bands have the Chern numbers $C=(-1,~+1,~0)$, respectively. (b) The corresponding strip spectrum for $g=0.4 $, with periodic boundary conditions applied along $\bm a_{1}=a_0(3,\sqrt{3})/2$. (c) Spatial distribution of the edge states, calculated as the sum of the probability of all localized modes within the bandgap. Parameters are set as $t=1$ and $\delta_\omega=3.2 $.}
\label{fig2}
\end{figure*}

\textit{Phonon-magnon model}---
We consider a honeycomb phononic crystal with one localized phonon mode on each A and B sublattice [Fig.~\ref{fig1}(b)]. A ferromagnetic island at each plaquette center supports a uniform Kittel mode \cite{Kittel1948,HolsteinPrimakoff1940,zhang2016cavity}.  The system is governed by the Hamiltonian
\begin{equation}
H=H_{\mathrm {ph}}+H_{\mathrm m}+H_{\mathrm{int}}.
\end{equation}
Here, $H_{\mathrm {ph}}$ is the phonon Hamiltonian given by
\begin{equation}
\begin{aligned}
H_{\mathrm{ph}} =& \sum_{r_{\mathrm A}} \omega_{\mathrm{ph}} a_{r_{\mathrm A}}^\dagger a_{r_{\mathrm A}} + \sum_{r_{\mathrm B}} \omega_{\mathrm{ph}} b_{r_{\mathrm B}}^\dagger b_{r_{\mathrm B}}\\
    &+\sum_{\langle r_{\mathrm A}, r_{\mathrm B} \rangle} t \left( a_{r_{\mathrm A}}^\dagger b_{r_{\mathrm B}} + \text{H.c.} \right),
\end{aligned}
\end{equation}
where $a_{r_{\mathrm{A}/\mathrm{B}}}^\dagger$ ($a_{r_{\mathrm{A}/\mathrm{B}}}$) and $b_{r_{\mathrm{A}/\mathrm{B}}}^\dagger$ ($b_{r_{\mathrm{A}/\mathrm{B}}}$) create (annihilate) a phonon mode at position $r_{\mathrm A}$ and $r_{\mathrm B}$, respectively, $\omega_{\mathrm {ph}}$ is the phonon frequency, $t$ is the nearest-neighbor phonon hopping, and H.c. stands for the Hermitian conjugate. Due to the large spatial separation between the ferromagnetic islands, the magnon Hamiltonian $H_{\mathrm m}$ only consists of on-site resonances:
\begin{equation}
H_{\mathrm m} =\sum_{R} \omega_{\mathrm m} m_{R}^\dagger m_{R},
\end{equation}
where $m_{R}^\dagger$ ($m_{R}$) are magnon creation (annihilation) operators at position $R$ and $\omega_{\mathrm m}$ is the magnon frequency. Without phonon-magnon interaction, the phononic crystal behaves like a phononic graphene with two Dirac points at the Brillouin corners (i.e., $\mathrm{K}$ and $\mathrm{K}'$ valleys), whereas the magnon dispersion is a flatband. This setup has the advantage that, once the Dirac points are properly gapped, a clean and full Chern gap can be easily obtained without further parameter optimization, and a low-energy description can directly capture the underlying mechanism.

The interaction term takes the form:
\begin{equation}
\begin{aligned}
H_{\mathrm{int}} = \sum_R\sum_{i=1}^{3}
\big[ &g\xi^{2(i-1)} a^\dagger_{R+\bm d_{\mathrm A,i}}m_R \\
&+g\xi^{2i-1} b^\dagger_{R+\bm d_{\mathrm B,i}}m_R +\mathrm{H.c.} \big].
\end{aligned}
\label{eq:Hint}
\end{equation}
Here, $g$ is the magnitude of the magnon--phonon coupling and $\xi=e^{-i\pi/3}$ is the phase increment between adjacent radial directions. This phase factor originates from projecting the right-circular Kittel mode onto the radial effective fields associated with the six surrounding phononic sites (see Supplemental Material (SM), Sec.~S1 for more details on the model construction~\cite{SM}). The A and B sites are labeled in their physical counterclockwise
order, with $\theta_{\mathrm A,i}=2\pi(i-1)/3$ and $\theta_{\mathrm B,i}=(2i-1)\pi/3$. For a fixed magnetization direction, these complex coupling phases generate the synthetic flux responsible for the Chern gap.

In the following, we consider a generic large-detuning scenario where $|\delta_\omega|\equiv |\omega_{\mathrm m}-\omega_{\mathrm{ph}}| \gg g, t$. In such a case, the magnon serves as an auxiliary degree of freedom to manipulate the phonon band topology, and the resulting topological transport remains phonon-dominated, as desired for phononic applications. In the Bloch basis $\Psi_{\bm k}=(a_{\bm k\mathrm A},a_{\bm k\mathrm B},m_{\bm k})^T$, the model Hamiltonian is~\cite{SM}
\begin{equation}
    H(\bm k)   =
    \begin{pmatrix}
    0 & t f(\bm k) & G_{\mathrm A}(\bm k)\\
    t f^*(\bm k) & 0 & G_{\mathrm B}(\bm k)\\
    G_{\mathrm A}^*(\bm k) & G_{\mathrm B}^*(\bm k) & \delta_\omega
    \end{pmatrix}, 
    \label{eq5}
\end{equation}
where $\bm k=(k_x,k_y)$ is the momentum and  $f(\bm k)=\sum_{i=1}^3 e^{i\bm k\cdot\bm d_{\mathrm{A},i}}$. The magnon--phonon coupling terms are given by
\begin{equation}
\begin{aligned}
G_{\mathrm A}(\bm k)&=g\sum_{i=1}^{3}\xi^{2(i-1)}e^{-i\bm k\cdot\bm d_{\mathrm A,i}},
\\
G_{\mathrm B}(\bm k)&=g\sum_{i=1}^{3}\xi^{2i-1}e^{-i\bm k\cdot\bm d_{\mathrm B,i}}.
\end{aligned}
\end{equation}
At the two Dirac points, the phases sum constructively on one sublattice and destructively on the other: $|G_{\mathrm A}(\mathbf K)|^{2}=9g^{2},~ |G_{\mathrm B}(\mathbf K)|^{2}=0$; and $|G_{\mathrm A}(\mathbf K')|^{2}=0, ~|G_{\mathrm B}(\mathbf K')|^{2}=9g^{2}$. Consequently, the magnon couples exclusively to sublattice A at $\mathrm K$ and to sublattice B at $\mathrm K'$, as plotted in Fig.~\ref{fig1}(c). This valley--sublattice locking generates opposite Dirac masses at the two valleys ($m_{\mathrm K/\mathrm K'}=\mp9g^2/2\delta_\omega$), ensuring that their Berry-curvature contributions constructively add up to open a topological Chern gap (see SM Sec.~S2~\cite{SM}). Interestingly, this mechanism closely mimics the gap-opening process in off-resonantly driven graphene, where photon-assisted transitions similarly generate a valley-opposite Dirac mass~\cite{Oka2009,Kitagawa2011,usaj2014irradiated,RudnerLindner2020}.

\begin{figure*}
\centering
\includegraphics[width=\textwidth]{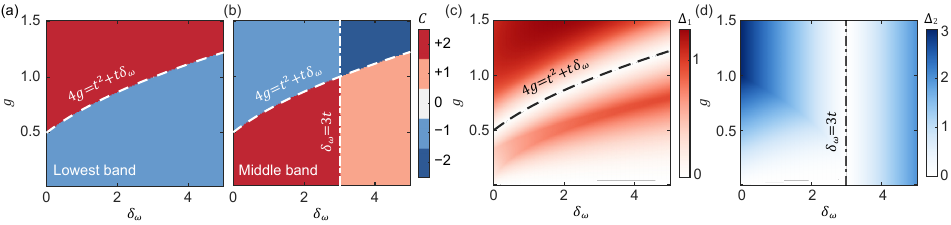}
\caption{
(a) and (b) Chern-number phase diagrams of the lowest band and the middle band, respectively, as functions of the magnon detuning $\delta_\omega$ and the magnon--phonon coupling strength $g$, with $t=1$. The dashed curve $4g^2=t^2+t\delta_\omega$, marks the gap-closing condition at the $\rm M$ point, while the vertical dashed line in (b) indicates the $\Gamma$-point transition at $\delta_\omega=3t$.
(c) and (d) Plots of bulk bandgaps $\Delta_{1}$ (gap between the lower and middle bands) and $\Delta_{2}$ (gap between the middle and upper bands), respectively, as functions of  $\delta_\omega$ and $g$.}
\label{fig3}
\end{figure*}

\textit{Effective Haldane-like model}---
To gain a deeper insight into the role of the central magnon on the phononic crystal, we apply the Schrieffer-Wolff transformation to derive the effective phonon Hamiltonian as~\cite{SchriefferWolff1966,Bravyi2011}:
\begin{equation}
    H_{\rm eff} = H_{\rm ph} - \sum_{R} \frac{V_R^\dagger V_R}{\delta_\omega},
\end{equation}
where  $V_R=\sum_{i=1}^{3}\left[ g\xi^{-2(i-1)} a_{R+\bm d_{\mathrm A,i}} + g\xi^{-(2i-1)} b_{R+\bm d_{\mathrm B,i}} \right]$. The magnon-phonon interaction generates four types of effective terms [see Fig.~\ref{fig1}(d)]: an on-site energy renormalization, nearest-neighbor (NN) hoppings, next-nearest-neighbor (NNN) hoppings, and third-nearest-neighbor (NNNN) hoppings. 
Specifically, the effective hopping amplitudes between two sites of the same sublattice around the same magnon center are
\begin{equation}
   t^{\mathrm{eff}}_{\mathrm A_i\leftarrow\mathrm A_j}= -\frac{g^2}{\delta_\omega}\, \xi^{2(i-j)}, \ \text{and}~
   t^{\mathrm{eff}}_{\mathrm B_i\leftarrow\mathrm B_j}=-\frac{g^2}{\delta_\omega}\,\xi^{2(i-j)},
\end{equation}
where $i,j$ label phonon sites surrounding the plaquette. For $i=j$, these expressions reduce to real on-site energy shifts $ -g^2/\delta_\omega$; for $i\neq j$, they yield complex NNN hoppings on the two sublattices. This is precisely the Haldane-type flux pattern that generates a nonzero Chern number without an external magnetic field~\cite{Haldane1988}. 

In addition, a magnon-mediated hopping from a $\mathrm B_j$ site to an $\mathrm A_i$ site carries the phase associated with the difference between their physical radial angles. It is given by
\begin{equation}
 t^{\mathrm{eff}}_{\mathrm A_i\leftarrow\mathrm B_j}=-\frac{g^2}{\delta_\omega}\,\xi^{2(i-j)-1}.
\end{equation}
This cross-sublattice term includes both NN and NNNN contributions, depending on the relative positions of the two sites.

\textit{Topological bandgap and boundary modes}---
To elucidate the emergence of the topological phase, we investigate the bulk band structure of the full coupled system. At $g=0$ the two phononic bands host gapless Dirac points at the $\rm K$ and $\rm K'$ valleys, while the magnon band is decoupled [Fig.~\ref{fig2}(a), dashed curves]. The chiral magnon--phonon coupling hybridizes these modes and opens a bandgap at the phonon Dirac points. Through a numerical calculation of the Chern number~\cite{fukui2005chern}, we find $C=\mp1$ for the two phonon-dominated bands and $C=0$ for the magnon-dominated bands [Fig.~\ref{fig2}(a), solid curves], consistent with previous analysis.

The nontrivial bulk topology implies the existence of chiral edge states via the bulk-boundary correspondence~\cite{thouless1982quantized}. For a strip geometry, Fig.~\ref{fig2}(b) shows the edge modes traversing the bulk gap, strictly localized at opposite boundaries and propagating unidirectionally. Figure~\ref{fig2}(c) further visualizes the boundary localization through the spatial intensity obtained by summing the probability of all in-gap states on a finite lattice. Reversing the magnetization direction ($+z\rightarrow -z$) complex-conjugates the local coupling phases $(\xi\rightarrow\xi^\ast)$, reversing the effective synthetic flux and changing the lowest-band Chern number from $-1$ to $+1$. Accordingly, the chiral edge states reverse their propagation direction.

\textit{Phase diagram and gap visibility}---
Finally, we work out the full phase diagram of the model by going beyond the large detuning limit and also allowing for strong magnon-phonon interaction. Figures~\ref{fig3}(a) and (b) show the evolution of the lower- and middle-band Chern numbers as functions of $\delta_\omega$ and $g$, with the dashed line denoting analytic phase boundaries. For the lower band, the Chern number changes from $-1$ to $+2$ at $4g^2 = t^2 + t\delta_\omega$, suggesting a large Chern number phase with multiple chiral edge states can be obtained via increasing the magnon-phonon hopping. In the large detuning limit, this phase transition can be inferred from the effective NNNN hopping [Fig.~\ref{fig1}(d)]. For the middle band, an additional phase transition occurs at $\delta_\omega=3t$, making the phase diagram much richer than the lowest band. Notably, when the sum of the Chern numbers of the lowest two bands is nonzero, the second bandgap will also host a chiral edge state~\cite{SM}.

Apart from the Chern numbers, gap size is also a crucial quantity that determines the performance of the system. Under strong detuning and weak coupling, the first bandgap $\Delta_{1}$ can be enlarged by reducing the frequency detuning $\delta_\omega$ or increasing the coupling $g$ [Fig.~\ref{fig3}(c)], consistent with the low-energy theory (i.e., $\Delta_{1}\propto g^2/\delta_\omega$). Away from this regime, the gap size is jointly determined by valley and $\rm M$ point modes, thus showing a nonmonotonic behavior. For the second bandgap, its size $\Delta_{2}$ vanishes along $\delta_\omega=3t$, which corresponds to the band touching between the middle and upper bands. The gap size generally grows away from the closing points.

A realistic GHz phononic-crystal platform can be implemented using SAW devices or suspended elastic crystals on LiNbO$_3$, AlN, or Si, with nearest-neighbor hopping rates of $t/2\pi\sim10$--$100~\mathrm{MHz}$ and phonon linewidths approaching $\gamma_{\rm ph}/2\pi\sim1~\mathrm{MHz}$ \cite{Shao2019,Zhang2022,Hatanaka2023,zhang2024gigahertz,Shen2020High,Zheng2025}. A CoFeB or YIG island placed at each plaquette center provides a Kittel mode whose frequency can be tuned by an external magnetic field, thereby enabling control of the magnon--phonon detuning $\delta_\omega$, while magnetoelastic and magnetorotational interactions generate the required coupling \cite{Kittel1958,Xu2020,Dreher2012,Liao2023}. CoFeB can offer relatively strong coupling, $g/2\pi\sim10$--$30~\mathrm{MHz}$~\cite{Hwang2024}, but its larger Gilbert damping can lead to magnon linewidths of $\gamma_{\rm m}/2\pi\sim10$--$100~\mathrm{MHz}$~\cite{Matsumoto2024}. YIG provides a favorable low-loss alternative, with $g/2\pi\sim5$--$10~\mathrm{MHz}$ and magnon linewidths near $\gamma_{\rm m}/2\pi\sim1~\mathrm{MHz}$~\cite{Kunstle2025}. For representative parameters $t/2\pi=30~\mathrm{MHz}$, $\delta_\omega=3.2t$, and $g/2\pi=7~\mathrm{MHz}$, the estimated topological gap is approximately $5~\mathrm{MHz}$, exceeding the expected phonon and YIG magnon linewidths and therefore allowing clear spectral resolution of the gap, which can be further enhanced by reducing the frequency detuning.

\textit{Conclusion---}
We have proposed a magnon-induced phononic Chern insulator in a graphene-type phononic crystal system. The key to the design is the placement of the phonon and magnon sites, which generates favored time-reversal-breaking hoppings and opens a full topological gap. The model is studied analytically by analyzing the full and reduced phonon Hamiltonians, which clearly reveal the core function of the magnon--phonon interaction. We also investigate the model numerically, showing the existence of the chiral edge states and the fruitful phase diagram.  More broadly, this work exemplifies an emerging human--AI discovery workflow in which the Qiushi Discovery Engine constructed the initial model and carried out the subsequent theoretical exploration and numerical checks, the authors then examined its physical assumptions, verified the results, and formulated the final interpretation.
These results demonstrate the power of using discrete lattices to engineer magnon-phonon interactions and establish magnons as tunable mediators of synthetic gauge fields for reconfigurable GHz phonon transport. 

\textit{Acknowledgements---}
This work was supported by the National Natural Science Foundation of China under Grant Nos. 62401491 (H.X.) 62175215 (Y.Y.), and U25D8017 (Y.Y.), the National Key Research and Development Program of China under Grant Nos. 2022YFA1405200 (Y.Y.), and 2022YFA1404900 (Y.Y.), the Class D for Young Scientific Research Peak Creation No. K20250230 (Y.Y.), the Fundamental Research Funds for the Zhejiang Provincial Universities No. 226-2025-00231 (Y.Y.), the Science Challenge Project N0. TZ2025015 (Y.Y.), the Research Grants Council of the Hong Kong SAR, China, under Grant No. 24304825 (H.X.), the Guangdong Provincial Quantum Science Strategic Initiative under Grant No. GDZX2501012 (H.X.), and the Chinese University of Hong Kong under Grant Nos. 4053729 (H.X.) and 4053794 (H.X.).

\bibliography{refs}

\end{document}